\documentclass[useAMS]{gGAF2e}

\usepackage{lineno}
\usepackage{psfrag}
\usepackage{bm}

\def\dsize{\displaystyle}
\def\E{{\rm E}}
\def\Ro{{\rm Ro}}
\def\Rm{{\rm R_m}}
\def\Re{{\rm Re}}
\def\Ra{{\rm Ra}}

\def\Pr{{\rm Pr}}
\def\q{{\rm q}}

\begin{document}
\doi{10.1080/03091920xxxxxxxxx}
 \issn{1029-0419} \issnp{0309-1929} \jvol{00} \jnum{00} \jyear{2010} 

\markboth{P.Hejda and M.Reshetnyak}{Nonlinearity in a dynamo}

\title{{\textit{Nonlinearity in a dynamo}}
}


\author{P. Hejda${\dag}$ 
\vspace{6pt} and M.Reshetnyak${\ddag}$ $^{\ast}$\thanks{$^\ast$Corresponding author. Email: m.reshetnyak@gmail.com}\\\vspace{6pt}  ${\dag}$Institute of Geophysics of the ASCR, Bo\v cn\'\i\;  II/1401,
141 31 Prague, Czech Republic\\ ${\ddag}$Institute of the Physics of the Earth RAS, B. Gruzinskaya 10, 
  Moscow,  Russia\\\vspace{6pt}\received{v3.3 released February 2009} }

\maketitle

\begin{abstract}
Using a rotating flat layer heated from below as an example, we  consider effects which lead to stabilizing an exponentially growing magnetic field in magnetostrophic convection in   transition from the kinematic dynamo to the full non-linear dynamo. We present estimates of the energy redistribution over the spectrum and helicity quenching by the magnetic field. We also study the alignment of the velocity and magnetic fields. These regimes are similar to those in planetary dynamo simulations.\bigskip
\end{abstract}

\bigskip

\section{Introduction}
Many physical processes can be referred to threshold phenomena, when the increase of the governing parameter 
 leads to the appearance of an increasing solution. Such an example is thermal convection, when 
 the growth of thermal instability  starts at a certain critical value of the Rayleigh number   $\rm Ra^{\rm cr}$, which characterizes the
 amplitude of the heat sources \citep{Chandra1961}. 
  The same situation occurs if magnetic field $\bf B$ is generatied in a conductive medium: 
  the increase of the convection intensity, parametrized by the magnetic Reynolds number ${\rm R_m}$, 
   can lead to an exponentially growing solution \citep{Moffatt1978}.
 After that magnetic field grows up to the moment when it starts to have an effect on the flow. 
  As in many astrophysical objects $\rm R_m$ is very large, providing quite extended spectra of the fields,     it  is generally believed  
    that this influence need not lead to a direct suppression of fluid motions.  This statement is supported by the fact that in some cases transition from the non-magnetic to the magnetic state can be accompanied by the growth of Reynolds numbers. In other words, knowledge of ${\rm R_m}$ is not sufficient to answer the question whether 
   the magnetic field will grow further or not.
   
The most widespread  point of view is that the magnetic field causes such a  reconstruction of the flow that the generation of the magnetic field becomes less efficient. However,   the visual control of the flow 
 does not reveal an essential change  \citep{Jones2000},  which can be due to the 
   force-free nature of the magnetic field configurations: $\Big{|}\left( \nabla\times {\bf B}\right)\times {\bf B} \Big{|} {\rm L_B}/B^2 \ll 1,$ where  $\rm L_B$  is the energy-carrying scale of the magnetic field. 
   
One of the explanations of the saturation mechanism is offered in \citep{Berger1984, BrSubr2005}, where saturation is connected with  the scale separation of the generated magnetic field due to conservation of magnetic helicity. In the present paper we study, on an example 
 of the dynamo in a rapidly rotating flat layer heated from below, how such a 
 magnetic energy redistribution over the spectra takes place in geostrophic systems. Our simulations demonstrate the occurrence of the magnetic $\alpha^M$-effect which can suppress the total $\alpha$-effect and lead to a saturated dynamo.  
 
 The other point is the correlation of velocity  and magnetic fields (so-called cross helicity). It appears that a saturated velocity field can still lead to an exponentially growing magnetic field, provided that this new artificial magnetic field does not contribute to the Lorentz force. It was first noticed by
 \cite{  CatTob2009}, see also their KITP's conference  video presentation. 
 This problem has now become the subject 
 of various discussions \citep{Tilgner2008, TBr2008, Schrinner2009}. We also present some results concerned with the magnetostrophic regimes close to those in geodynamo simulations that are also unstable for large $\Rm$.  
 
 \section{Dynamo equations}
The geodynamo equations for an incompressible fluid 
  ($\nabla\cdot{\bf V}=0$) in a layer of height  $\rm L$
 rotating with angular velocity $\Omega$  in a Cartesian system of
  coordinates $(x,\,y,\,z)$
 in its traditional geodynamo  dimensionless form can
be expressed as follows:
\begin{equation}
\begin{array}{l}\dsize
   \frac {{\partial}{\bf B}}{ {\partial t}} ={ \nabla}\times \left({\bf V}\times{\bf B}\right)+
\q^{-1} \Delta {\bf B} \\ \\
\dsize
    \E\Pr^{-1}\left[\frac {{\partial} {\bf V}} {\partial t}+ \left({\bf
V}\cdot \nabla\right) {\bf V}\right] = -\nabla { P}  -{\bf
{1}_z}\times{\bf V} + \\ \\  \qquad\qquad \Ra { T} \,z{\bf{1}_z}+
\left(
\nabla\times {\bf B}
\right)\times {\bf B}+
 \E\Delta {\bf V}
\\  \\ \dsize
\frac{\partial { T}} {\partial t}+\left({\bf V}\cdot\nabla\right)
\left({ T}+{ T}_0\right)= \Delta{T}.
\end{array}\label{sys0}
\end{equation}
Velocity $\bf V$, magnetic field $\bf B$, pressure
$P$ and the typical
diffusion time $t$ are measured in units of
$ \rm \kappa/L$,   $\dsize \sqrt{2\Omega\kappa\mu\rho};$, $\rm
\rho\kappa^2/L^2$ and $\rm L^2/\kappa$, respectively, where $\kappa$ is
the thermal  diffusivity, $\rho$ is the density, $\mu$ the permeability,
$\dsize \Pr=\frac{\kappa}{
\nu}$ is the Prandtl number,
  $ \dsize \E =\rm  \frac{\nu}{ 2\Omega L^2}$ is the
Ekman number, $\nu$ is the kinematic viscosity,
$\eta$ is the magnetic diffusivity, and ${\q}=\kappa/\eta$ is the
Roberts number.
$\dsize \Ra\rm
=\frac{\alpha g_0\delta T {L}}{  2\Omega\kappa}$ is the modified
Rayleigh number, $\alpha$ is the coefficient of volume expansion,
$\delta T$ is the unit of temperature,
 $g_0$ is the gravitational acceleration, and $T_0=1-z$ is the heating from below.
 The problem is closed with periodical boundary conditions in the $(x,\, y)$ plane. In the $z$-direction, we use
  simplified conditions  \citep{Cattaneo2003}: for $T$ and $\bf V$: 
$T=0$ (heating from below), stress-free for $\bf V$: 
  $\dsize V_z={\partial V_x\over \partial z}=
 {\partial V_y\over \partial z}=0$,  and the pseudo-vacuum boundary condition for $\bf B$: $\dsize B_x=B_y={\partial B_z\over \partial z}=0$ at $z=0,\,1$.
 System (\ref{sys0}) was solved using the pseudo-spectral Fortran-95 MPI code  \citep{RH2008} on cluster PC   computers using grids $N^3,\, N=64$.

\section{Results of modelling} \label{p41}
\subsection{General properties}
 We consider two regimes that differ in Rayleigh and Roberts numbers: 
\begin{enumerate}
 \item[R1:] $\Ra=4\cdot 10^2$, $\Pr=1$, $\E=2\cdot 10^{-5}$, $\rm q=10$.
 \item[R2:] $\Ra=1\cdot 10^3$, $\Pr=1$, $\E=2\cdot 10^{-5}$, $\rm q=3$.
\end{enumerate}

\begin{figure}[th!] \vskip -2cm
\center{
\epsfig{figure=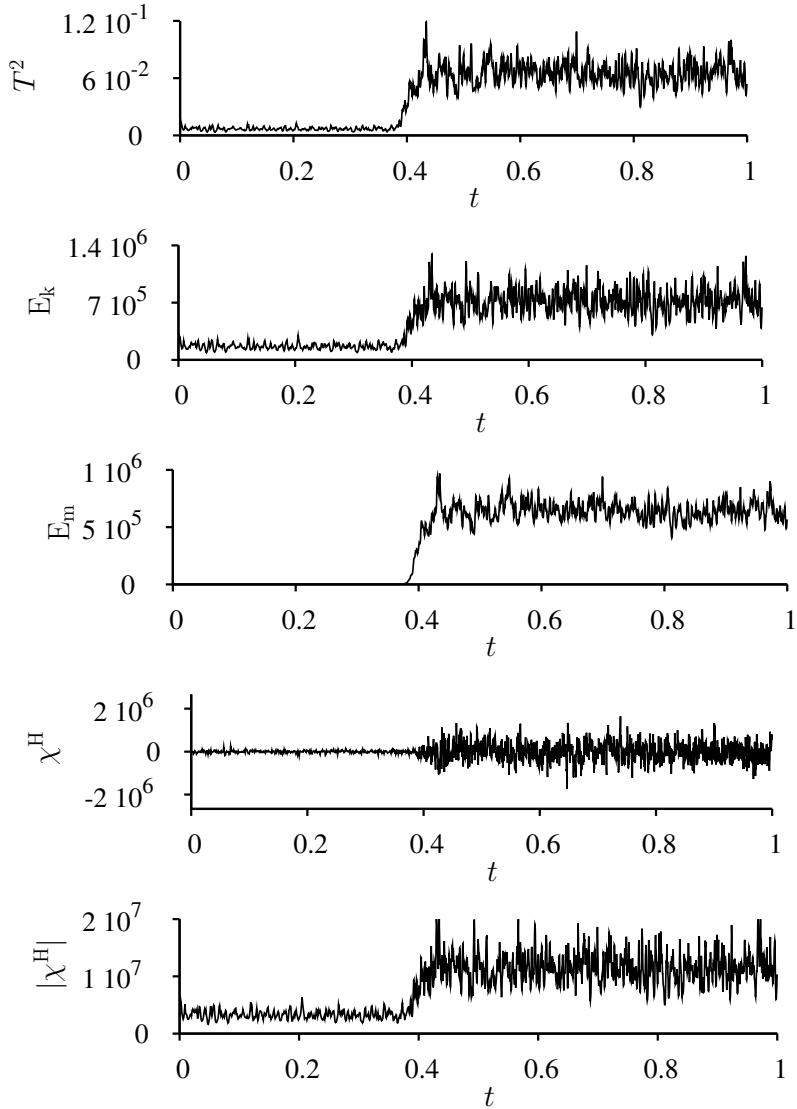,width=14cm}}
\vskip -3cm
 \caption{ Regime R1. 
 Evolution of the mean over the volume fluctuations of the squared temperature 
   $T^2$, kinetic and magnetic energies 
    $E_K$, $E_M$, and kinetic helicity  $\chi^{ H}$, $|\chi^{ H}|$. 
} \label{Fig1}
\end{figure}

\begin{figure}[th!] \vskip -1cm
\center{\epsfig{figure=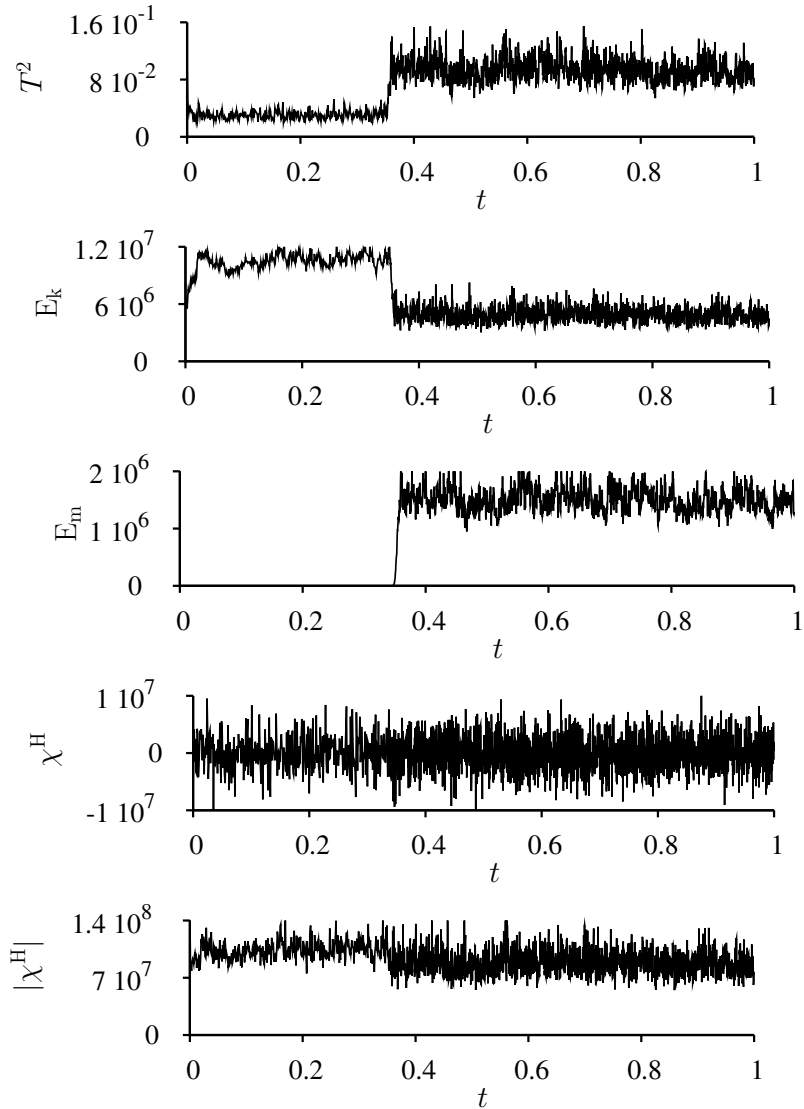,width=14cm}}
\vskip -3cm
\caption{The same as in Fig.~\ref{Fig2} for regime R2.
} \label{Fig2}
\end{figure}

To get the final saturated dynamo, we started from a pure convection state without the magnetic field. At time $t=0.35$, we injected a magnetic field of small amplitude.
 After an intermediate kinematic regime with exponential growth of magnetic energy $\dsize E_M={B^2\over 2\Ro}$, we arrive at a quasi-periodical state with  kinetic, $\dsize E_K={V^2\over 2}$, and magnetic, $E_M$, energies of the same order of magnitude for R1 ( see Fig.~\ref{Fig1}), and at the state with $E_K>E_M$ for R2 (see  Fig.~\ref{Fig2}), where a smaller value of $\q$ was used.  

\begin{figure}[th!]
\vskip 0cm\center{
\begin{minipage}[h!]{.45\linewidth}
\vspace*{2mm}
\vskip -0cm
\vskip -0cm     
\hskip 1cm \includegraphics[width=8cm]{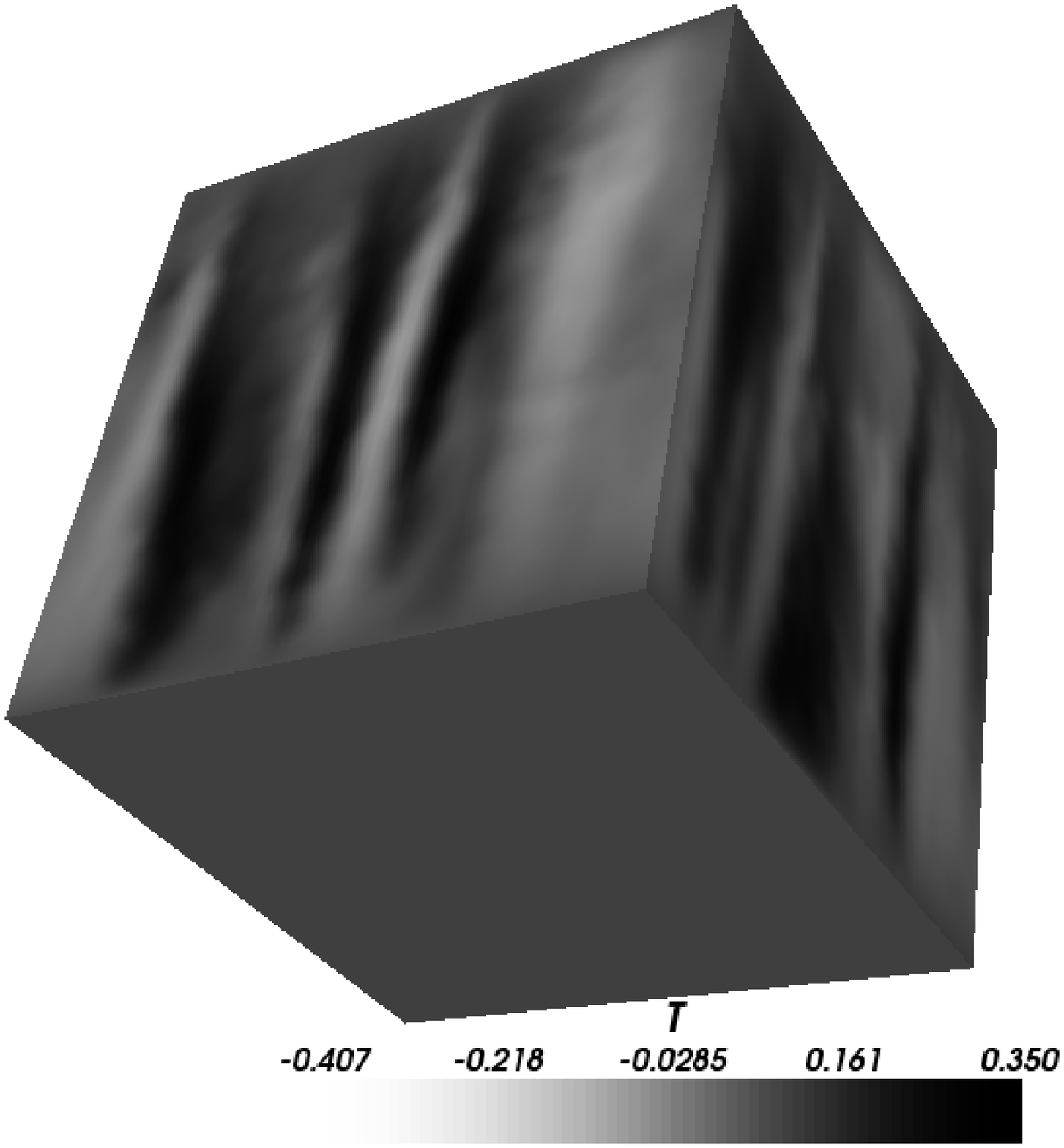}
\end{minipage}\hfill
\begin{minipage}[h!]{.45\linewidth}
\vspace*{2mm}
\vskip -0cm
\vskip -0cm  
 \vskip 0cm
\hskip -2cm \includegraphics[width=8cm]{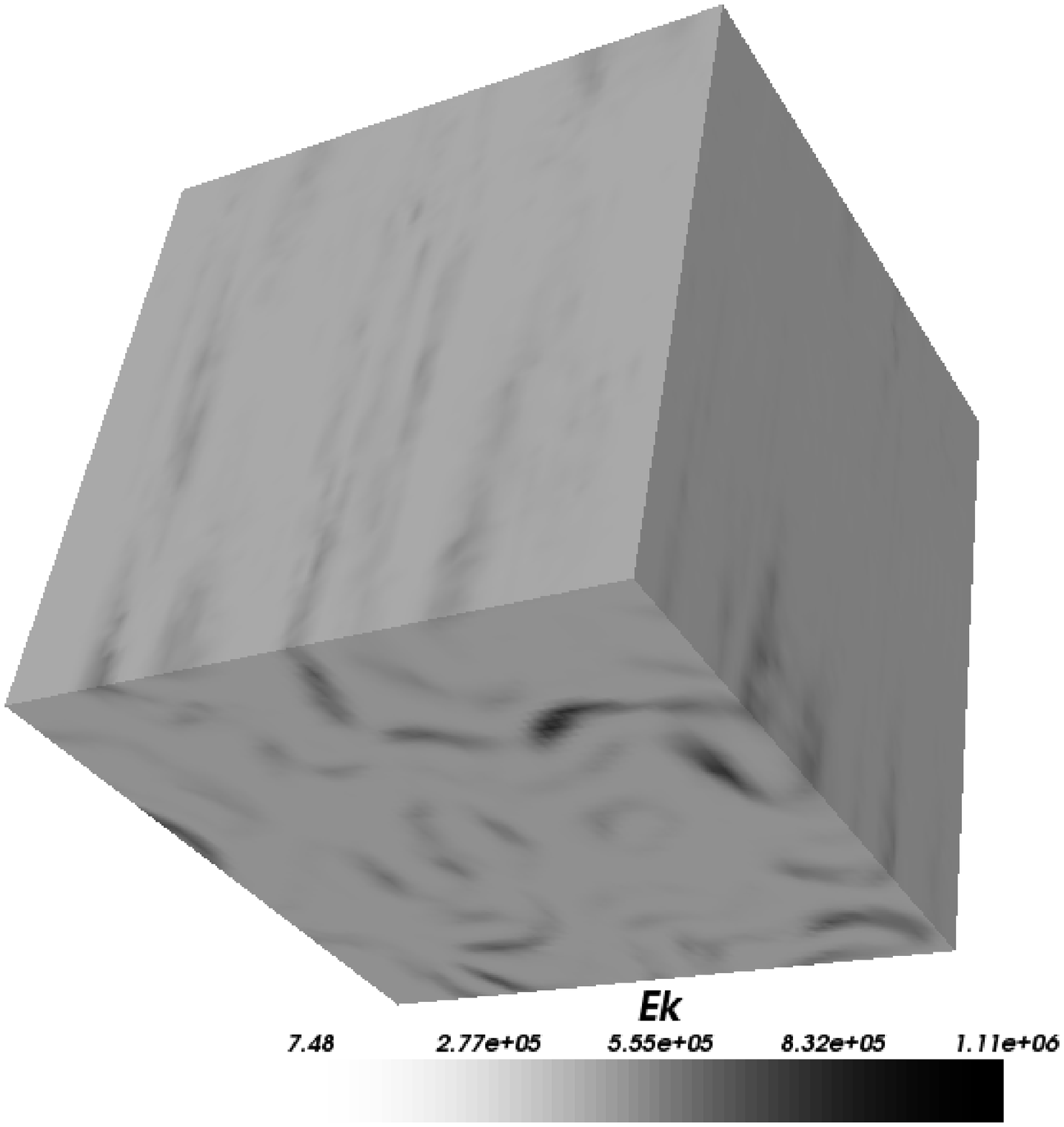}
\end{minipage}}\hfill\vskip 1cm 
 \caption{
 Distribution of temperature fluctuations  $T$ and kinetic energy
   $E_K$ for regime R2.  
} \label{Fig3}
\end{figure}

\pagestyle{empty}
Both regimes correspond to the geostrophic (magnetostrophic) state, see typical cyclonic structures
 of the temperature fluctuations  $T$ and  kinetic energy $\dsize E_K={V^2\over 2}$   in Fig.~\ref{Fig3} for R2.
 The diameter of the cyclones depends on the Ekman number as $d_c\sim {\rm L}\E^{1/3}$ \citep{Busse1970}.
 In both cases the transition from the kinematic regime to the full dynamo was accompanied by an increase of the temperature fluctuations    $T^2$. 

 On the other hand, the behaviour of the kinetic energy for the two regimes was different: for R1 the growth of the magnetic field leads 
  to  an increase in convection intensity ($E_K$ becomes larger) and for R2 $E_K$ to a decrease. 
 The corresponding Reynolds numbers are $\Re=200$ (kinematic dynamo regime) and $\Re=420$ (regime with saturation)
   for R1 and $\Re=1500,\, 1100$ for R2.   For the observed geostrophic state the ratio of the nonlinear term in the Navier-Stokes equation to the Coriolis force, the Rossby number,      $\Ro=\E V$ is quite small:  $\Ro=4.3\, 10^{-3},\, 8.4\, 10^{-3}$ for R1 and $\Ro=3.0\, 10^{-2},\, 2.2\, 10^{-2}$ for R2.      One of the explanations of the growth of $\Re$ for regime R1 is that the magnetic field disturbs the regular geostrophic flow for R1   and helps the generation of the magnetic field. The total energy $E_K+E_M$  also increases. 
 Regime R2 corresponds    to the more disturbed state, and  the kinetic energy is reduced to the amount of 
  increase of the magnetic energy. This scenario usually takes place for large Reynolds numbers. Note also that, while for R1 the increase of $T^2$ corresponds to the increase of the Archimedean work $\Ra T V_z$,  for R2 the work decreases due to the 
  chaotization of the fields. The behaviour of the kinetic helicity  $\dsize \chi^{ H}={\bf V}\cdot {\rm rot }{\bf V}$    as a whole copies the evolution of the kinetic energy. The dispersion of $\dsize \chi^{ H}$ increases when the magnetic field    reaches its saturated value.
\pagestyle{empty}
\subsection{Spectra}
\begin{figure*}[t]
\vskip 4cm
\psfrag{k}{\large $k$}
\begin{minipage}[h!]{.45\linewidth}
\vspace*{2mm}
\vskip -5cm
\psfrag{V}{ ${\bf V}{\bf V}^*$}
\vskip -0cm      \hskip 7cm  1 
\hskip 1cm \includegraphics[width=8cm]{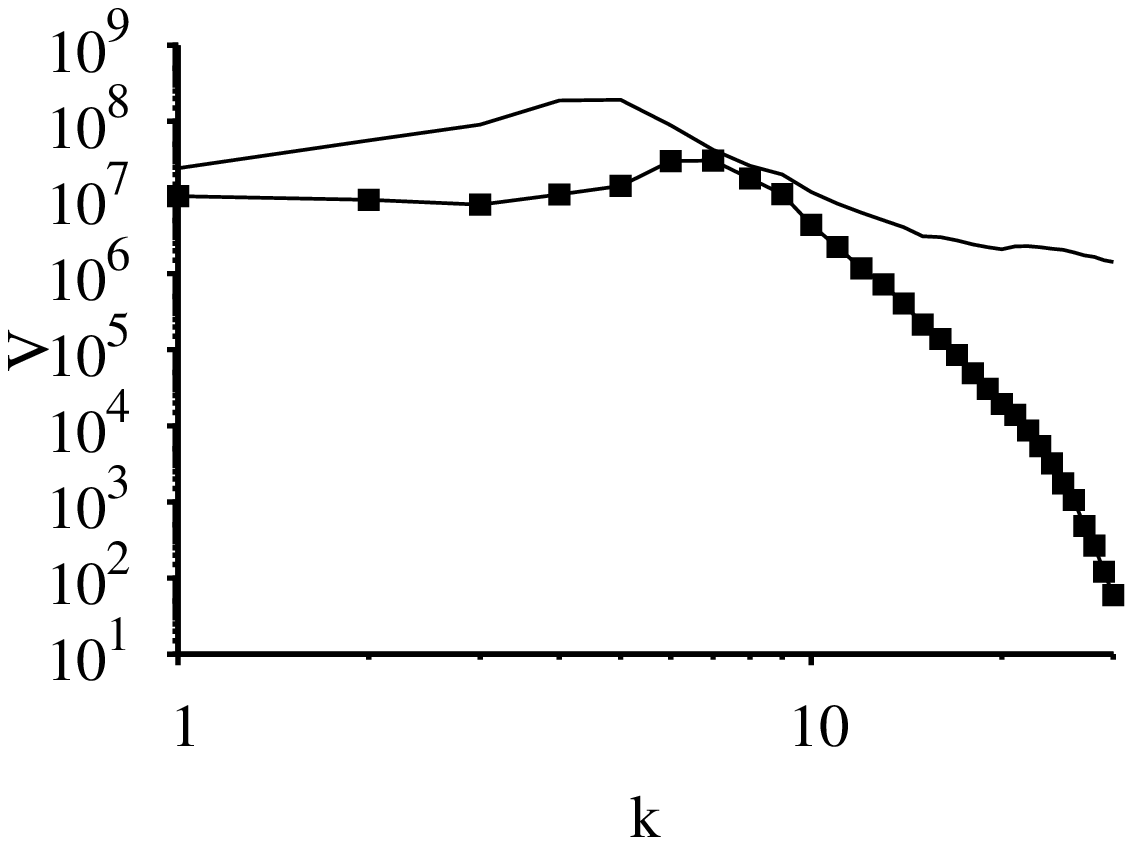}
\end{minipage}\hfill
\begin{minipage}[h!]{.45\linewidth}
\vspace*{2mm}
\vskip -5cm
\psfrag{V}{ ${\bf B}{\bf B}^*/{\rm Ro}$}
\vskip -0cm \hskip 6cm  2 
 \vskip 0cm
\hskip 0cm \includegraphics[width=8cm]{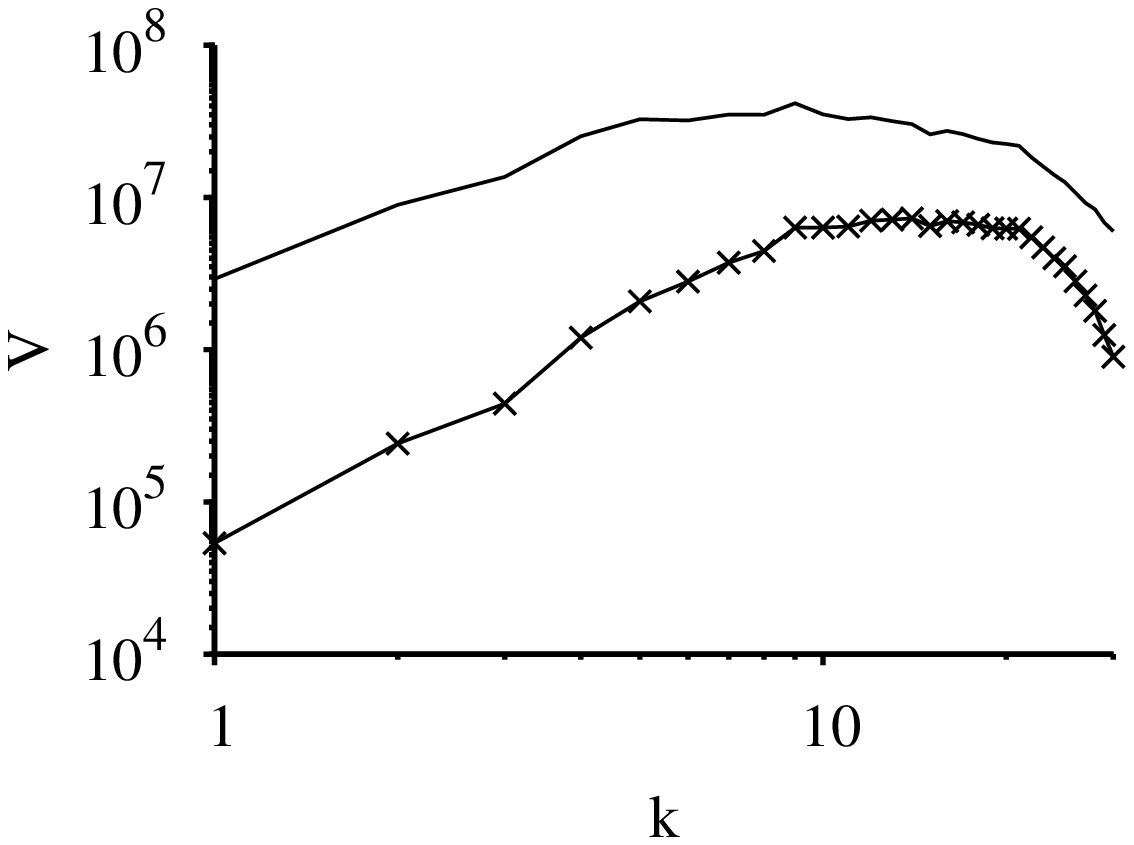}
\end{minipage}\hfill
\vskip 1cm
\begin{minipage}[h!]{.45\linewidth}
\vspace*{2mm}
\vskip -0cm
\psfrag{V}{ ${\bf V}{\bf V}^*$}
\vskip -0cm      \hskip 7cm  3 
\hskip 1cm \includegraphics[width=8cm]{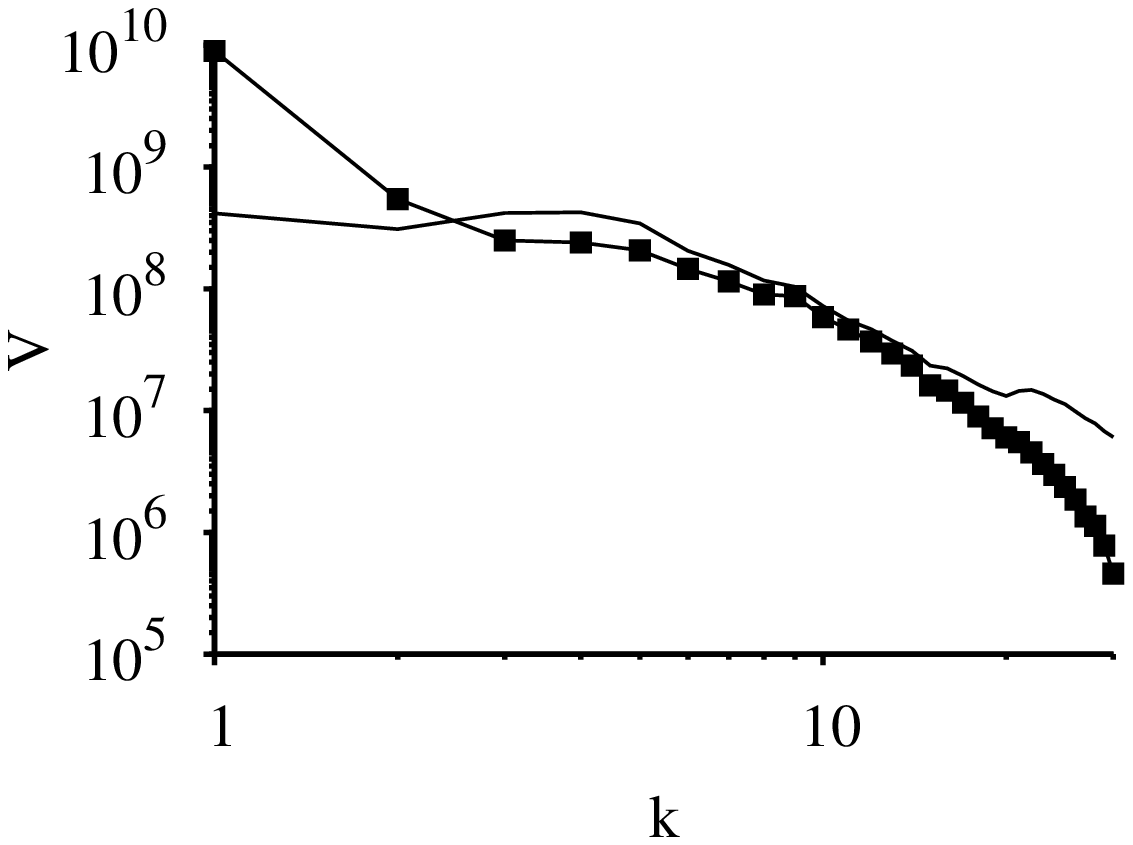}
\end{minipage}\hfill
\begin{minipage}[h!]{.45\linewidth}
\vspace*{2mm}
\vskip -0cm
\psfrag{V}{ ${\bf B}{\bf B}^*/{\rm Ro}$}
\vskip -0cm \hskip 6cm  4 
 \vskip 0cm
\hskip 0cm \includegraphics[width=8cm]{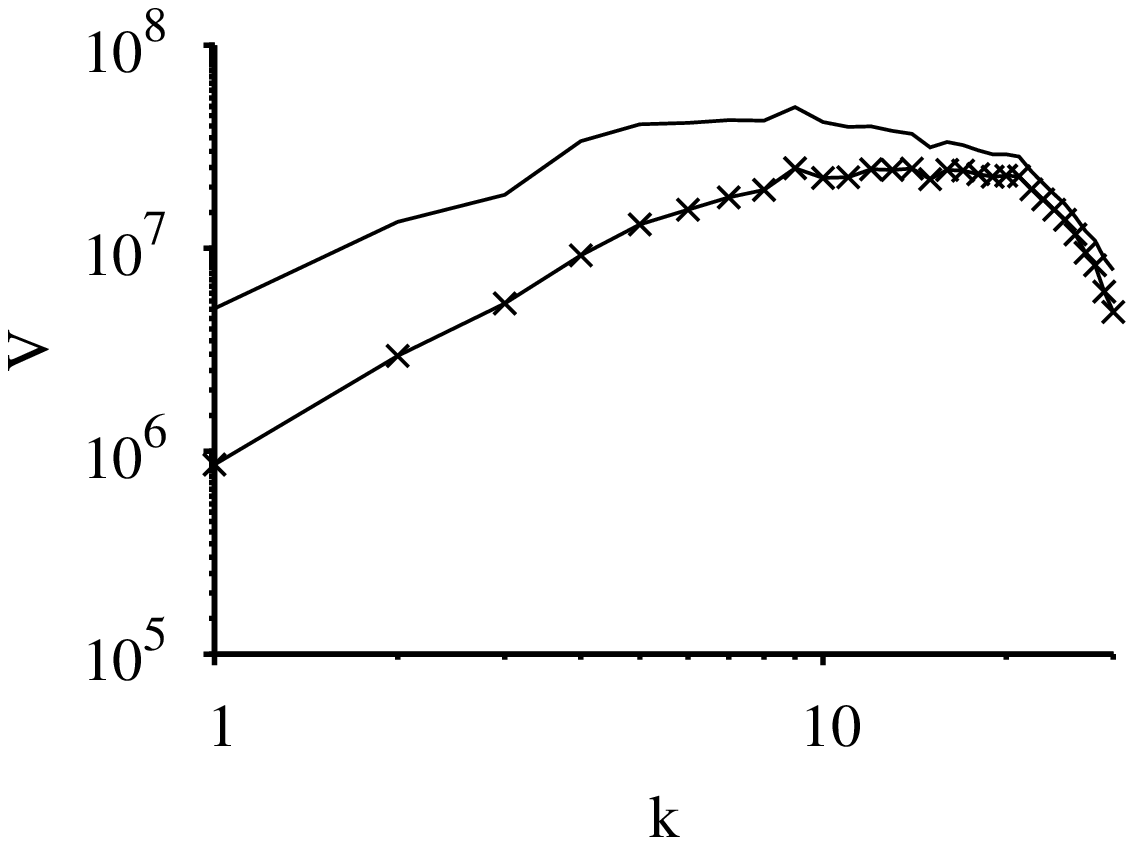}
\end{minipage}\hfill
\vskip 0cm 
\caption{
Spectra of the kinetic and magnetic energies for regime R1 (1-2) and R2 (3-4).
 The solid line correspond to the saturated dynamo, squares to the convection without magnetic field, and crosses to the kinematic dynamo regime. 
  }\label{Fig4}
\end{figure*}

 The maximum of the kinetic energy spectra for R1 corresponds to the horizontal scale of the cyclones
  $k_c=1/d_c$, see Fig.~\ref{Fig4}(1). The magnetic field slightly decreases this maximum and kinetic energy increases. The relative part of the kinetic energy on the large scales also increases.
 
The growth of the magnetic field during transition from the kinematic dynamo state to  saturation state 
 is accompanied by an increase of the magnetic field on large scales Fig.~\ref{Fig4}(2).
  The first mode which reaches saturation is the mode with  $k\sim 10$. The other modes still grow filling the spectra for $E_M$. This behaviour is the same for both the regimes R1, R2. The behaviour of the kinetic energy on the large scales is a little bit different for R2 ( Fig.~\ref{Fig4}(3)), where we observe a decrease of $E_K$, which corresponds to the breakup of the horizontal rolls by the magnetic field. The maximum for $E_K$ at    $k=1$ then disappears. 
  
The inhomogeneous growth of the magnetic field for different $k$ is quite important for understanding the saturation mechanism of the magnetic field. The growth of the magnetic field in the kinematic state takes place  at convective times  $\tau_k\sim (k V_k)^{-1}$, which decreases with $k$. According to \cite{Kazantsev1968} 
  spectrum of the magnetic field is  $E_M\sim k^{3/2}$ for the  non-rotating turbulence, and the maximum of the magnetic field is then close to the dissipative scales.   
  In our regime, the maximum of   $E_K$ at  $k=k_c$ is more important 
 for the magnetic field distribution, and the modes with  $k\sim k_c$ reach saturation level at first. We argue that this regime is closely connected with the occurrence of the  coherent structures discussed in \citep{TobCat2008}. 

\pagestyle{empty}

The observed redistribution of the magnetic field over the scales is closely related to the mean over the volume 
 magnetic helicity 
 $	 {\chi}^{\cal M}={\bf A}\cdot {\bf B}$  for $\Rm\to \infty$ \citep{Berger1984}, where 
$\bf A$ is the vector potential of the magnetic field  ${\bf B}={\rm rot }{\bf A}$\footnote{Large $\rm R_m$ is typical for many astrophysical bodies which possess their own magnetic field.}. 
 The alternating-sign quantity $ {\chi}^{\cal M}$ is a measure of the linkage of the magnetic field lines with one another \citep{Moffatt1978}. After some algebra with the induction equation, one arrives at the equation for the 
  mean over the volume fields \citep{Berger1984, BrSubr2005}
 \begin{equation}\begin{array}{l}\dsize
{D \over D t }\overline{\bf A\cdot B}=-{\rm R_m^{-1}}  \overline{\bf J\cdot B} +\Pi,
\end{array}\label{sys1}
\end{equation} 
where $ {\bf J}= {\rm rot }{\bf B}$ is the current,  
 and  $\Pi$ is the flux of  ${\chi}^{\cal M}$  through 
 the boundary. $\Pi=0$ for the fully periodical boundary conditions as well as for the super conductive boundaries  . 
 The other scalar  product $\dsize {\chi}^{\cal J}={\bf J}\cdot {\bf B}$ is the so-called current helicity. 

Then, after time  $t\sim \Rm\gg 1$, one has saturation regime $\dsize {D \over D t }=0$, with 
\begin{equation}\begin{array}{l}\dsize
{\bf \overline{J}\cdot \overline{B}}=-\overline{\bf j\cdot b}, \hskip 1cm
{\bf \overline{A}\cdot \overline{B}}=-\overline{\bf a\cdot b},
\end{array}\label{sys1}
\end{equation}
where the decomposition of field $F$ into the mean and fluctuating parts was used: $F=\overline{F}+f$ \citep{BrSubr2005}.  
 In other words, after the kinematic regime no changes of the magnetic field would change the total magnetic and current helicities. Any local change is possible only due to the redistribution of $ {\chi}^{\cal M}$, $ {\chi}^{\cal J}$ over the scales. 
Of course, this approach becomes more complicated, if the mean helicities change sign in space, and the ideas 
 on scale decomposition are applicable to the space domain with the same sign of helicity. 

The application of the pseudo-vacuum boundary conditions, which provides   $\Pi\ne 0$ at the boundaries,  is more tricky.
 These conditions leads to the increase of the energy of the magnetic field at large scales \citep{BrSand2004} and break the
  catastrophic quenching predicted by \cite{VainCat1992} and observed for the fully periodic boundary conditions, see  \citep{Hughes1996, BrSubr2005}. There is also an indication of catastrophic quenching for the mean-field dynamo models with periodic boundary conditions, see \citep{Cattaneo1996}.  Accordingly to this scenario  the large-scale magnetic field $B_0$ would be saturated 
  at $B_0^2\sim {\rm R_m^{-1}} b^2$, where $b^2\sim v^2$ are the small-scale field energies.

\begin{figure}[th]
\large
\psfrag{V}{\LARGE $|{\bf B}{\,{\rm rot} \bf B}^*|\over {\rm Ro}$}
\psfrag{k}{\large $k$}
\center{
\hskip 0.0cm\epsfig{figure=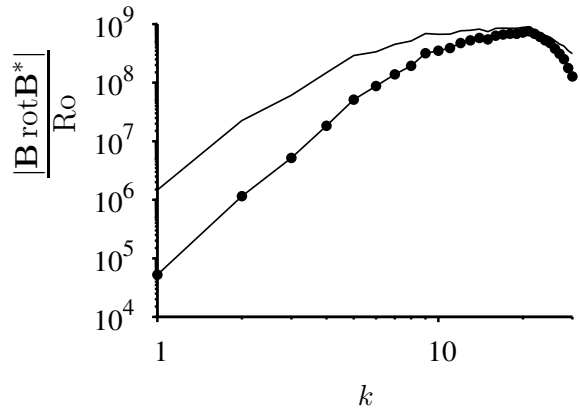,width=8cm} }
\vskip 0cm
\caption{Spectra of the current helicity for regime R1. The solid line corresponds to 
 the saturated dynamo, circles to the kinematic dynamo regime. }\label{Fig5}
\end{figure}

In our simulations with  pseudo-vacuum boundary conditions, we still observe some decrease of the magnetic field intensity, see Fig.~\ref{Fig1}--\ref{Fig2}: the ratio of the magnetic to kinetic energies decreases  from R1 to R2, which should be explained by a decrease of the Roberts number $\q$ rather  than by catastrophic quenching. 
 In Fig.~\ref{Fig5} we also observe that the  magnetic helicity on the large scales increases sufficiently after transition to the saturated regime  .

\begin{figure*}[t]
\vskip 0cm
\psfrag{z}{\large $z$}
\begin{minipage}[h!]{.45\linewidth}
\vspace*{2mm}
\psfrag{k1}{ }
\psfrag{a1}{1}
\hskip 1cm \includegraphics[width=7cm]{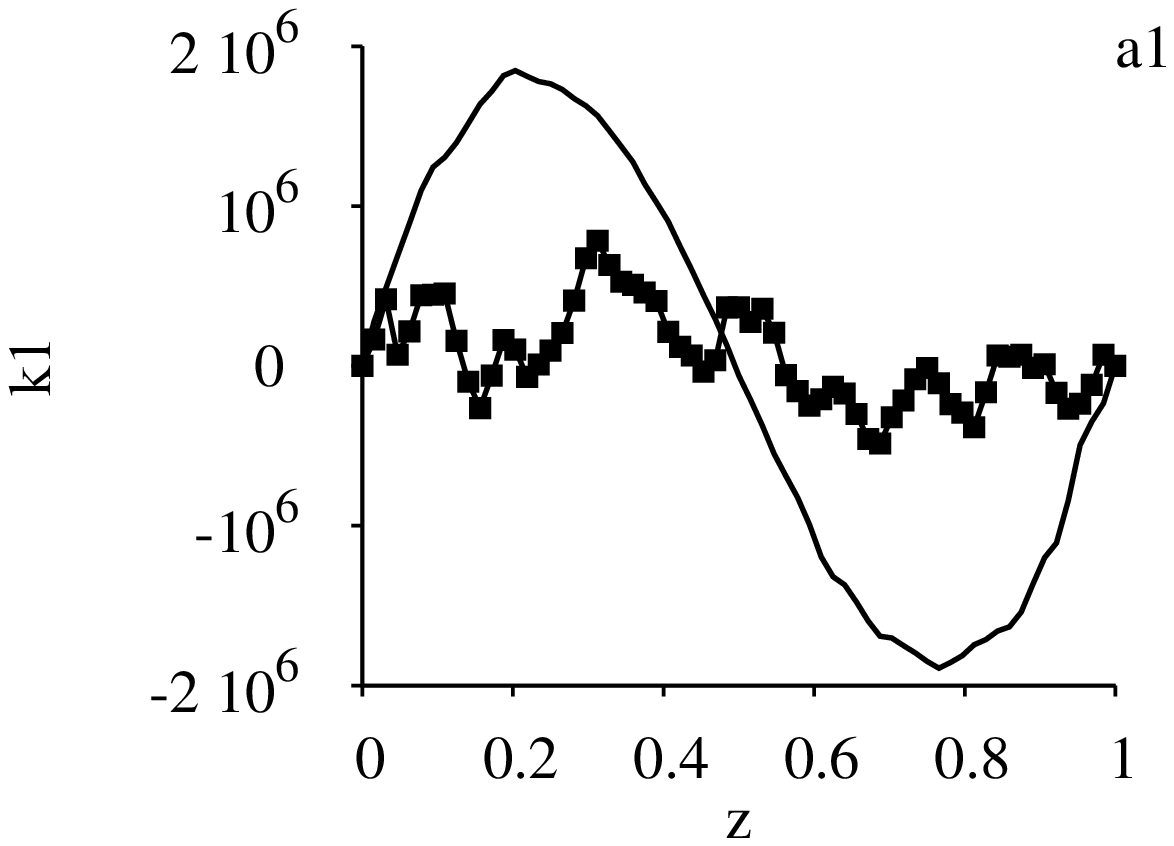}
\end{minipage}\hfill
\begin{minipage}[h!]{.45\linewidth}
\vspace*{2mm}
\vskip -0cm
\psfrag{k1}{ } \psfrag{a1}{2}
\hskip -1cm \includegraphics[width=7cm]{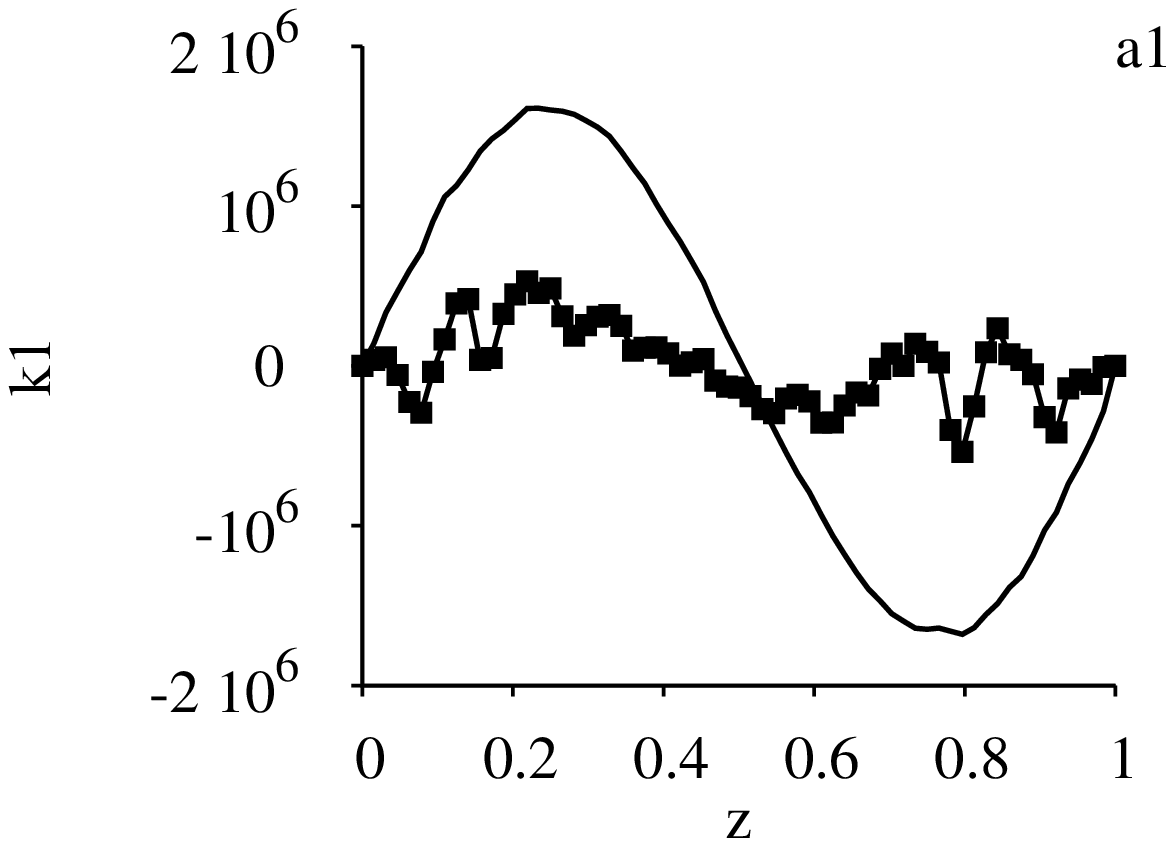}
\end{minipage}\hfill
\vskip 0.5cm
\caption{ Z-profiles of the kinetic   $\chi^{ H}$ (solid line) and 
  current  $\chi^{\cal J}$ (squares) helicities for regimes  R1 (1) and R2 (2). }\label{Fig6}
\end{figure*}

\begin{figure*}[t]
\psfrag{z}{\large $z$}
\begin{minipage}[h!]{.45\linewidth}
\psfrag{k1}{   $\chi^{\cal M}$} \psfrag{a1}{1}
\hskip 1cm \includegraphics[width=7cm]{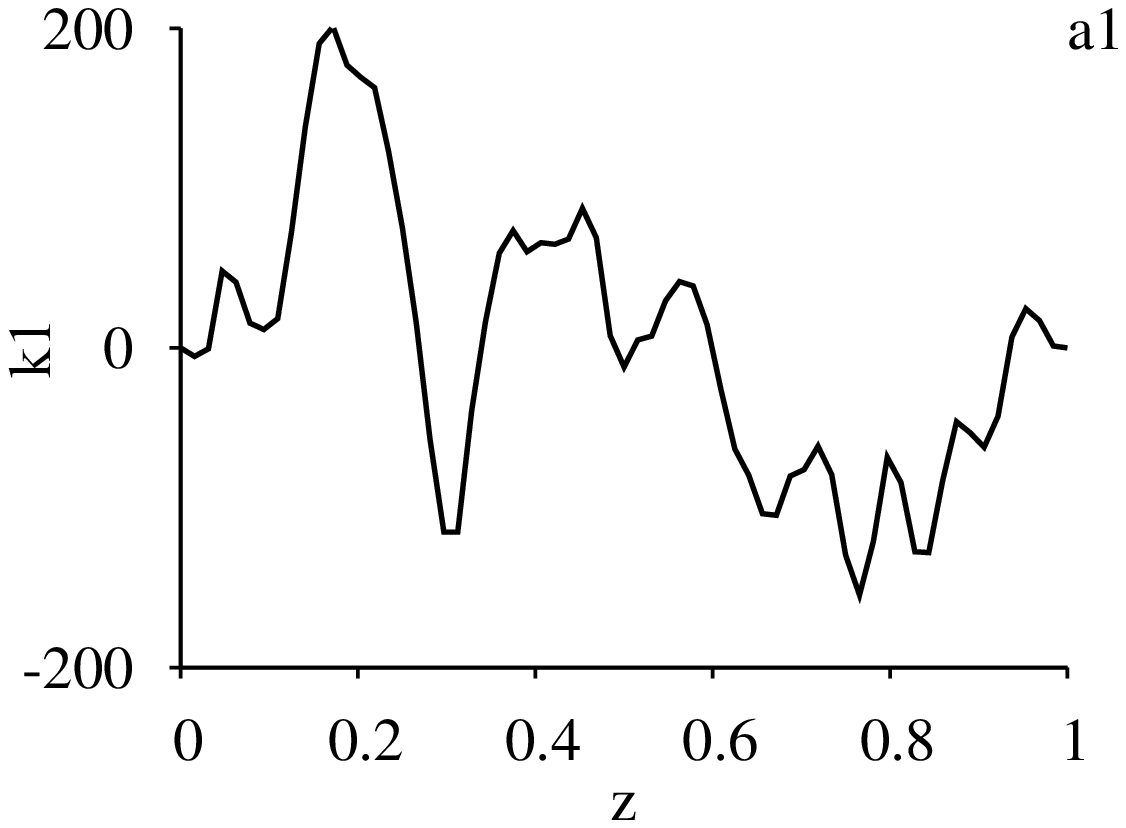}
\end{minipage}\hfill
\begin{minipage}[h!]{.45\linewidth}
\vskip -0cm
\psfrag{k1}{   $\chi^{\cal C}$ } \psfrag{a1}{2}
\hskip -1cm \includegraphics[width=7cm]{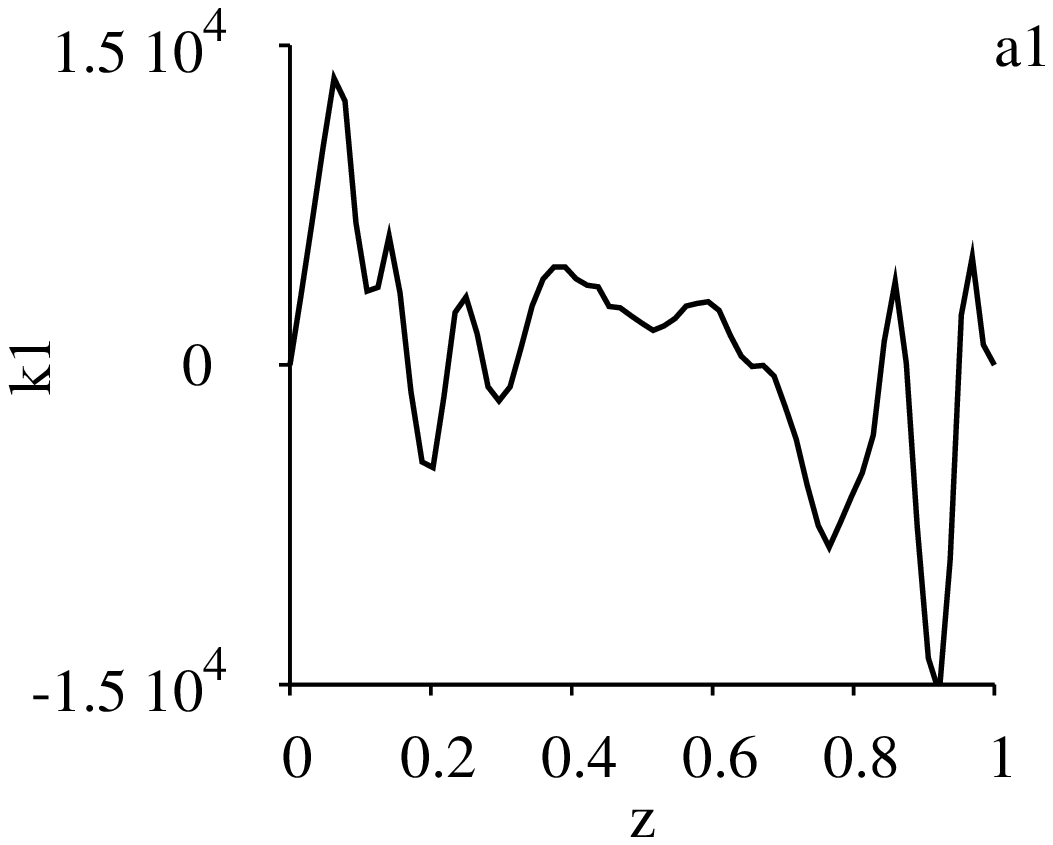}
\end{minipage}\hfill
\vskip -1cm \hskip 0cm   \vskip 1cm
\begin{minipage}[h!]{.45\linewidth}
\psfrag{k1}{    $\chi^{\cal M}$} \psfrag{a1}{3}
\hskip 1cm \includegraphics[width=7cm]{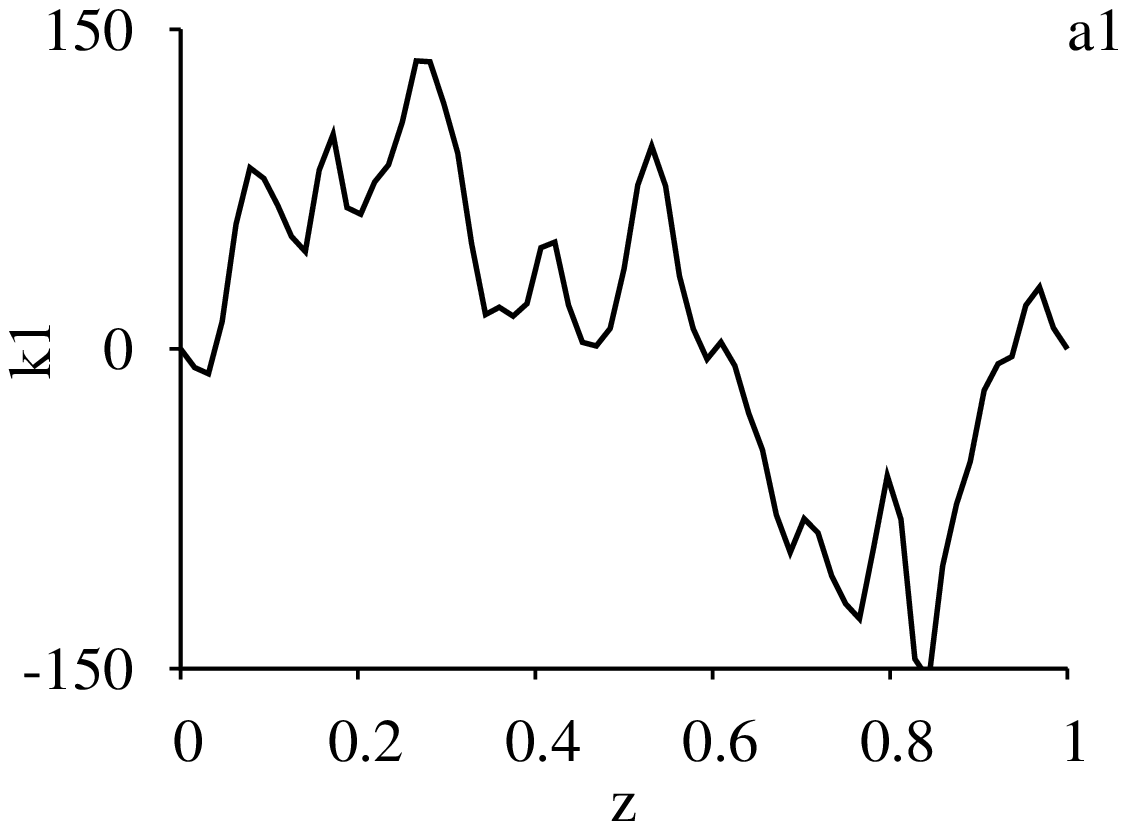}
\end{minipage}\hfill
\begin{minipage}[h!]{.45\linewidth}
\psfrag{k1}{    $\chi^{\cal C}$ } \psfrag{a1}{4}
\hskip -1cm \includegraphics[width=7cm]{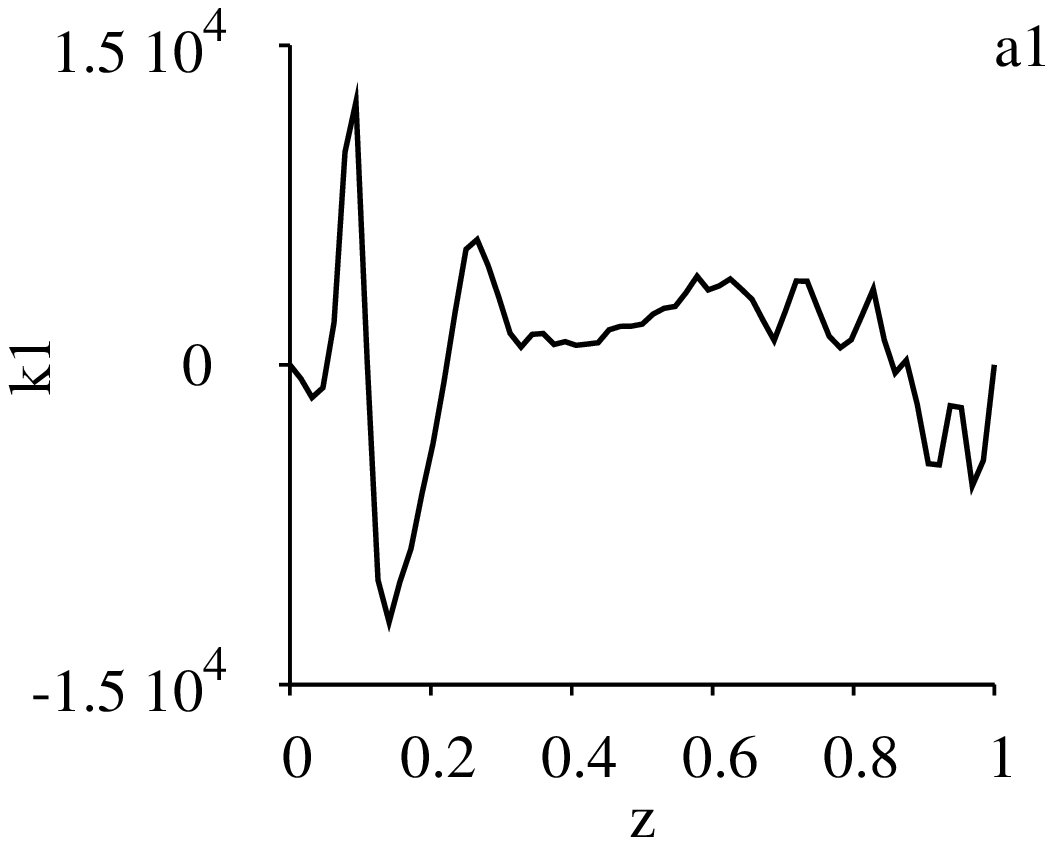}
\end{minipage}\hfill
\vskip 0.5cm
\caption{ Z-profiles of the magnetic $\chi^{\cal M}$ and cross-  $\chi^{\cal С}$ helicities for regime R1 (1-2) and R2 (3-4). }\label{Fig7}
\end{figure*}

Let us recall that, according to the mean field dynamo theory, there is connection between the
 kinetic and current helicities and hydrodynamic $\alpha^{\cal H}$-  and magnetic $\alpha^{\cal M}$-effects for short correlation times  \citep{Pouquet1976, Zeldovich1983} 
\begin{equation}\begin{array}{l}\dsize
\alpha^{\cal H}=-\tau \overline{{\bf v}\cdot {\bm \omega}}/3,\hskip 1cm 
 \alpha^{\cal M}= \tau \overline{{\bf j \cdot b}}/3,
\end{array}\label{sys2}
\end{equation}
where
${\bm \omega={\rm rot}{\bf v}}$ is the vorticity and  $\tau$ is a correlation time. In practice 
 \citep{Zeldovich1983}
 these formulas are applicable when the typical time of the large-scale magnetic field growth $\tau_{\rm L}$ is larger than the turnover kinetic time $\tau_t=l/ v(l)$, where $v(l)$ is a velocity on scale $l$. A rough estimate of $\tau_{\rm L}$, using jump of the magnetic energy at the kinematic regime  R1,  (see Fig.~\ref{Fig1}), yields $\tau_{\rm L}\sim 10^{-2}$, which is already smaller than $\tau_t\sim 1/\left(\Re k_c\right)\sim 10^{-3}$ (here we supposed that the kinetic energy is concentrated in the vicinity of $k_c$). 
 Taking into account that  the mean magnetic field grows  slower  than the small-scale field, we find that 
$\tau_{\rm L}$ is even much larger than the above estimate (a similar situation for regime R2 takes place). 

The total  $\alpha$-effect is then 
\begin{equation}\begin{array}{l}\dsize
\alpha=\alpha^{\cal H}+\alpha^{\cal M}.
\end{array}\label{sys3}
\end{equation}
If the signs of the helicities are the same, the total $\alpha$-effect is reduced  
($\alpha^{\cal M}\to -\alpha^{\cal H}$) and the magnetic field stops growing. 
 The latter is well observed in our simulations, Fig.~\ref{Fig6}--\ref{Fig7},
  where  $\chi^{\cal C\, J\,   M}> 0$ for $z<0.5$ and  $\chi^{\cal C\, J\,   M}< 0$ for  $z>0.5$. 
 All the three helicities $\chi^{\cal C\, J\,   M}$  have the same signs, however, $\chi^{\cal J}$ is more rugged due to the contribution of the small-scale fields.

\begin{figure*}[t]
\vskip -7cm
\center{
\hskip -3.0cm\epsfig{figure=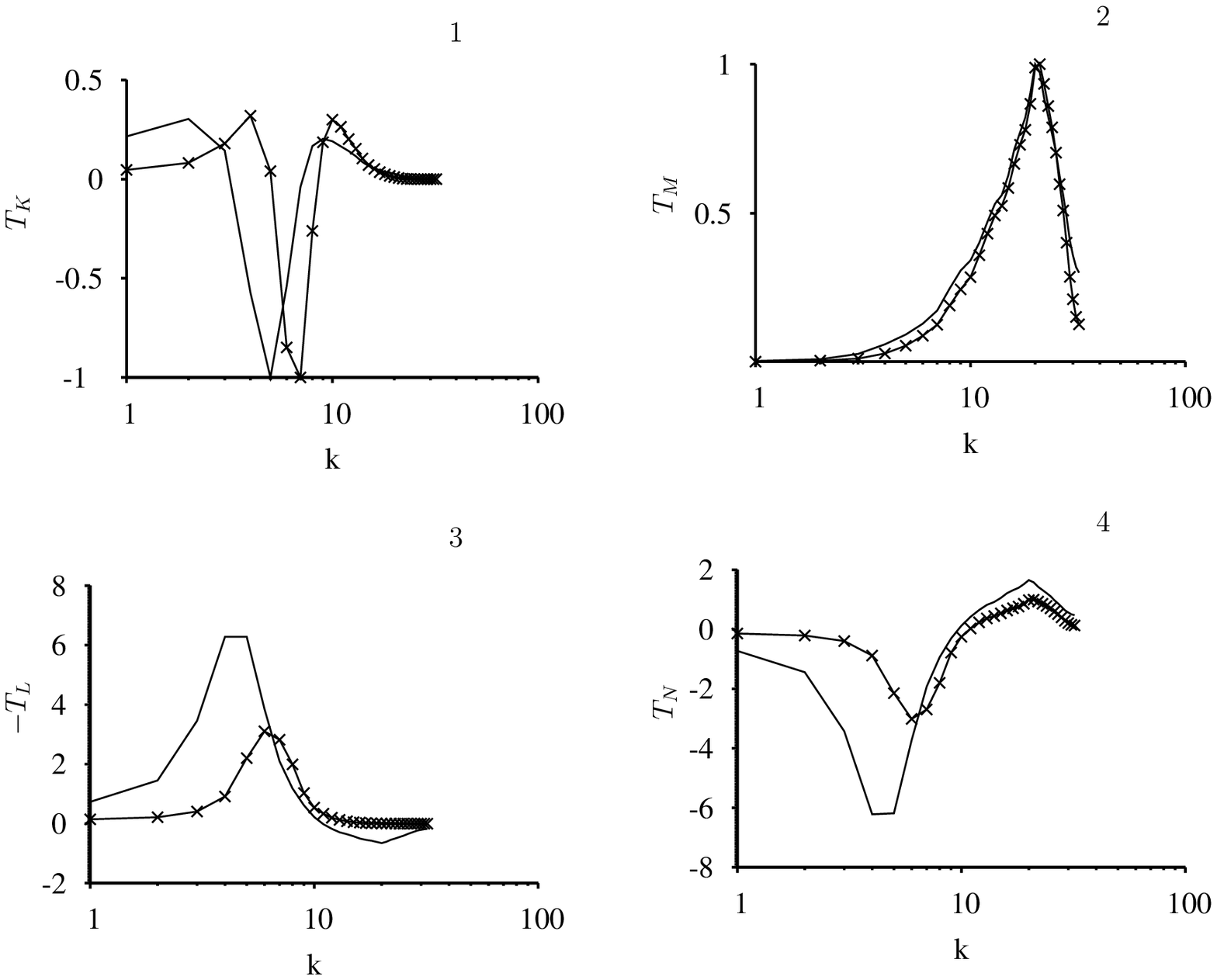,width=18cm}}
\vskip -7cm
\caption{Fluxes in the wave space of the kinetic  $T_K$ and magnetic  $T_M$ energies,
 Lorentz work  $T_L$ and the convective term in the induction equation  $T_N$,  regime  R1.
 The solid line corresponds to the saturated dynamo and crosses  to the kinematic dynamo regime.}\label{Fig8}
\end{figure*}

\subsection{Energy fluxes in the wave space}
Energy redistribution over the spectra is closely related to the fluxes in wave space. In spite of the fact that the final saturated state is quasi-stationary, there are still fluxes in wave space. This happens because the scales where energy is generated and dissipated are different \citep{RoseSulem1978}. For the simplest cases, such as 3D Kolmogorov's turbulence, kinetic energy goes from the scale of the force to the the dissipative scale. 
 Here we repeat the results of \citep{RH2008, HR2009} and discuss the difference between energy fluxes of kinematic dynamo regime and the saturated regime.
\begin{figure*}[t]
\vskip -7cm
\center{
\epsfig{figure=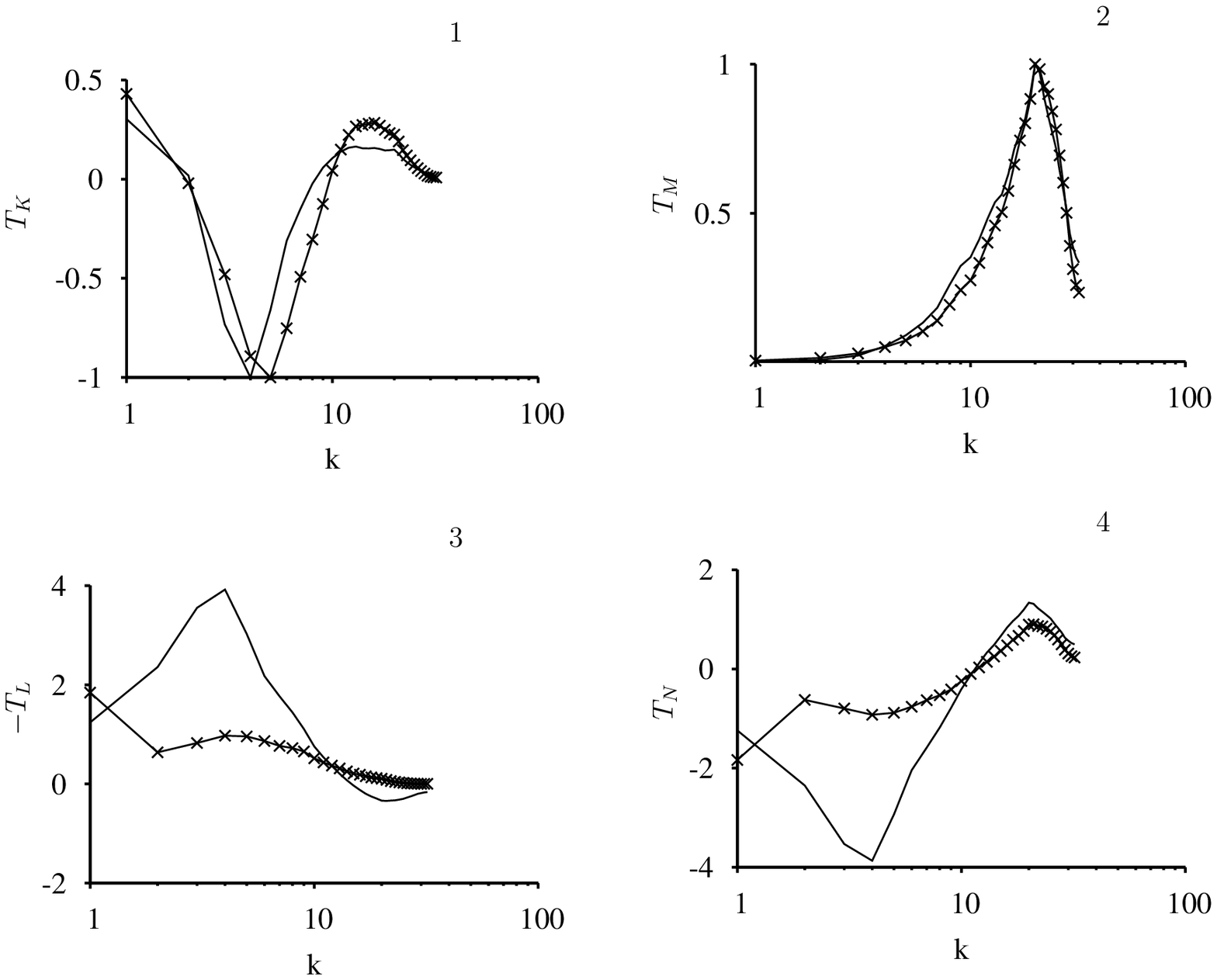,width=18cm}
}
\vskip -7cm
\caption{The same as in Fig.~\ref{Fig8} for regime R2.}\label{Fig9}
\end{figure*} 

 Consider the kinetic energy flux in wave space  through wave number  $k$: $\dsize {\cal T}_K(k)=-{\partial \over \partial k} 
<\left({\bf V}\times {\rm rot}{\bf V}\right)\cdot {\bf V}^<> 
  $, where 
${\bf V}^<({\bf r})=\sum\limits_{|{\rm P}|\le k} \widehat{{\bf V}}_{\rm P}\,e^{i{\bf P}{\bf r}}$  is a low frequency counterpart, and 
   $
\int\limits_{k=0}^{\infty} {\cal T}_K(k)\, dk=0 
$. 

Using relation 
${\rm rot}\left({\bf V}\times {\bf B}\right)=-\left({\bf V}\cdot \nabla \right) {\bf B}+
\left({\bf B}\cdot \nabla \right) {\bf V}$, the  magnetic energy flux  ${\cal T}_M(k)$ can be decomposed into advective and generating parts
 ${\cal T}_M={\cal T}_N-{\cal T}_L$, $\int\limits_{k=0}^{\infty} {\cal T}_N(k)\, dk=0$. 
 The latter is equal to minus the work of the Lorentz force.
 
The fluxes of kinetic energy ${\cal T}_K$ are presented  in Fig.~\ref{Fig8}(1), \ref{Fig9}(1).
 For $k\sim k_c$ the inverse cascade of the kinetic energy is observed: the cyclones are the sources of the energy. Here energy is 
  distributed to the large scales  
  ($k<k_c$, ${\cal T}_K>0$, inverse cascade), as well as to the small scales ($k>k_c$, ${\cal T}_K<0$, direct cascade) where it dissipates. The magnetic field causes  some blurring of the maxima and shift of 
  ${\cal T}_K$ to the large-scale region.

In contrast to  ${\cal T}_K$, ${\cal T}_M$ includes 
 not only an advective term, but also a generating term. It appears that the  integral  of ${\cal T}_M$ over 
 all $k$ is positive. Moreover,  
     ${\cal T}_M$ is positive for any $k$, Fig.~\ref{Fig8}(2), \ref{Fig9}(2), i.e. the magnetic field
      is generated on small scales. Position of the maximum of   ${\cal T}_M$ 
      is close to the maximum in the spectrum of  $E_M$.
 The form of ${\cal T}_M$ is the same for the small magnetic field during kinematic regime and for the saturated mode. 
 Let us consider where the  magnetic energy comes  from: Is the  energy transfered from other scales to some certain $k$,
  or is it  generated localy? Generating flux $-{\cal T}_L$ is shown  in Figs~\ref{Fig8}(3), \ref{Fig9}(3).  Its maximum is close to the minimum of  ${\cal T}_K$ (R1), 
   i.e the kinetic energy is transformed to the energy of the magnetic field.  Except for a small region at high $k$, 
    there is always an inverse cascade of magnetic energy $-{\cal T}_L>0$. 
 Note that the amplitudes of the maximum of     $-{\cal T}_L$ and ${\cal T}_M$ 
 are sufficiently different. That is because   $-{\cal T}_L$ and ${\cal T}_N$ are anti-correlated, see Fig.~\ref{Fig8}(4), \ref{Fig9}(4). This means that advective term ${\cal T}_N<0$ 
 transports the major part of  the energy to the dissipative scale. 
  
Transition to the saturated state for R1 does not change   form of the flux's curve substantially. It only increases the amplitudes of 
   $-{\cal T}_L$ and 
  ${\cal T}_N$, thus increasing the synchronization of the fluxes 
  $-{\cal T}_L$ and ${\cal T}_N$. The maxima of  $-{\cal T}_L$ and ${\cal T}_N$ shift to the large scales for the
   full dynamo regime.    
  For  the R2-regime saturation changes the flux of the magnetic energy more efficiently.
 For the kinematic regime, the flux was more or less homogeneous from the small to large scales. For the saturated state, 
 cyclones at $k_c$ conected with 
  decrease of  $E_M/E_K$ provide the main contribution  to $-{\cal T}_L$. The relative strength of the total flux   of the magnetic energy due to the $-{\cal T}_L$ term to the main scale remains at the same level. This flux can be related to the $\alpha$-effect.  

\section{Alignment of the fields}

In discussing z-profiles of the fields, we have not yet taken into account the importance of the spatial-temporal correlation of fields $\bf V$ and $\bf B$. In this connection an interesting question arises:  Can an already  quenched velocity field in the dynamo model  generate an exponentially growing magnetic field or not \citep{CatTob2009, TBr2008}? The only one difference between this new passive magnetic field and the original one is that the new field does not contribute to the Lorentz force.  It appears that the answer depends on the spectrum of the magnetic field and its time behaviour. Usually, if one has only one excited magnetic field mode for the saturated regime \citep{TBr2008} the new, passive magnetic field is stable. This statement is supported by simulations in the sphere \citep{Tilgner2008, Schrinner2009}, where the dipole mode dominates for low Rossby number $\rm R_o$ . For the multi-mode regime (which is the case for the large Rayleigh  number) the situation is different: the new magnetic field $\widehat{\bf B}$ grows exponentially \citep{CatTob2009}. There is an indication that the threshold for such a different behaviour of  $\widehat{\bf B}$ in time in the sphere takes place at  $\rm R_o=0.12$ \citep{Schrinner2009}. Here we consider what happens in the flat layer dynamo for regimes R1, R2 with quite extended spectra, adding a new induction equation for $\widehat{\bf B}$ to the system (\ref{sys0}):
   \begin{equation}\begin{array}{l}\dsize
{\partial {\bf B}\over\partial t}={\rm rot}\left({\bf V}\times {\bf B}\right)
+ {\rm q}^{-1}\Delta {\bf B} \\ \\
\dsize
    \E\Pr^{-1}\left[\frac {{\partial} {\bf V}} {\partial t}+ \left({\bf
V}\cdot \nabla\right) {\bf V}\right] = -\nabla { P}  -{\bf
{1}_z}\times{\bf V} + \\ \\  \qquad\qquad \Ra { T} \,z{\bf{1}_z}+
\left(
\nabla\times {\bf B}
\right)\times {\bf B}+
 \E\Delta {\bf V}
\\  \\ \dsize
{\partial { T}\over\partial t}+\left({\bf V}\cdot\nabla\right)
\left({ T}+{ T}_0\right)= \Delta { T} \\ \\ \dsize
{\partial {\bf \widehat{B}}\over\partial t}={\rm rot}\left({\bf V}\times {\bf \widehat{B}}\right)
+ {\rm q}^{-1}\Delta {\bf \widehat{B}}.
\end{array}\label{sys00}
\end{equation}
\begin{figure*}[t]
\vskip -1cm
\psfrag{t}{  $t$} \psfrag{k1}{$  E_m$}  \psfrag{a1}{1} 
\hskip 0.0cm\epsfig{figure=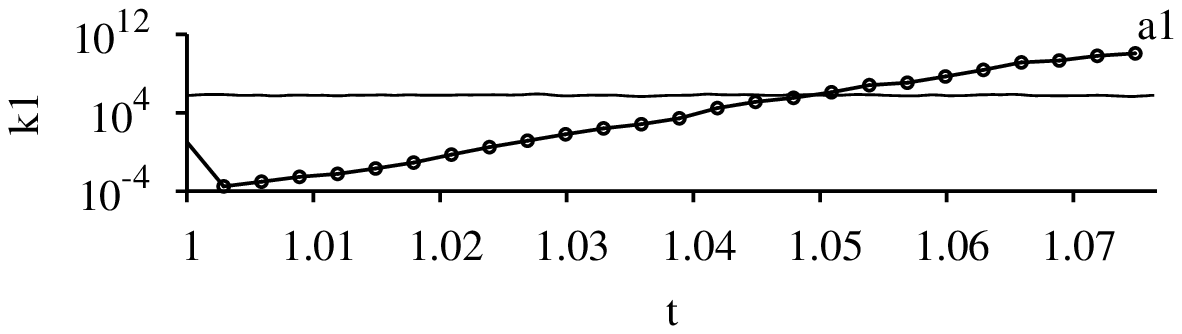,width=14cm}
\psfrag{t}{  $t$} \psfrag{k1}{$  E_m$}  \psfrag{a1}{2}
\hskip 0.0cm\epsfig{figure=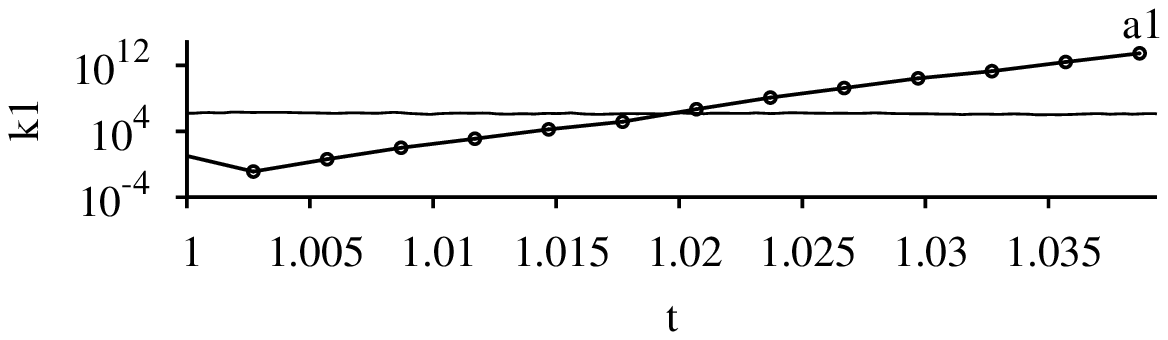,width=14cm}
\vskip 0cm
\caption{ Evolution of the magnetic energies  $B^2/2$ (solid line) and 
  $\widehat{B}^2/2$ (circles) for regimes  R1 (1) and R2(2).}\label{Fig10}
\end{figure*}
\begin{figure*}[t]
\vskip -6cm
\hskip 0.0cm\epsfig{figure=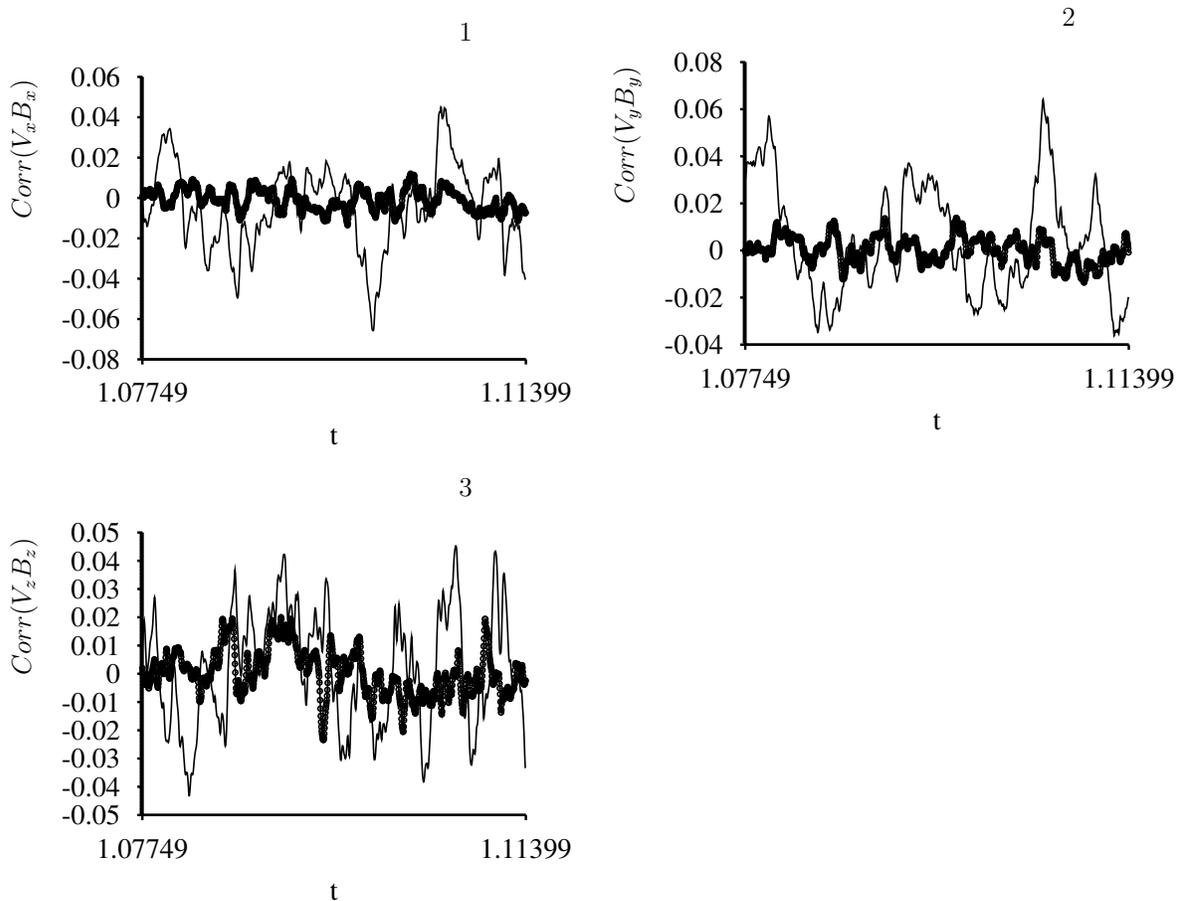,width=18cm}
\vskip -8cm
\caption{ Evolution of the autocorrelation functions for regime R1. The thin line corresponds to the original magnetic field $B$ and the line with circles to $\widehat{B}$.}\label{Fig11}
\end{figure*}

\begin{figure*}[t]
\hskip 0.0cm\epsfig{figure=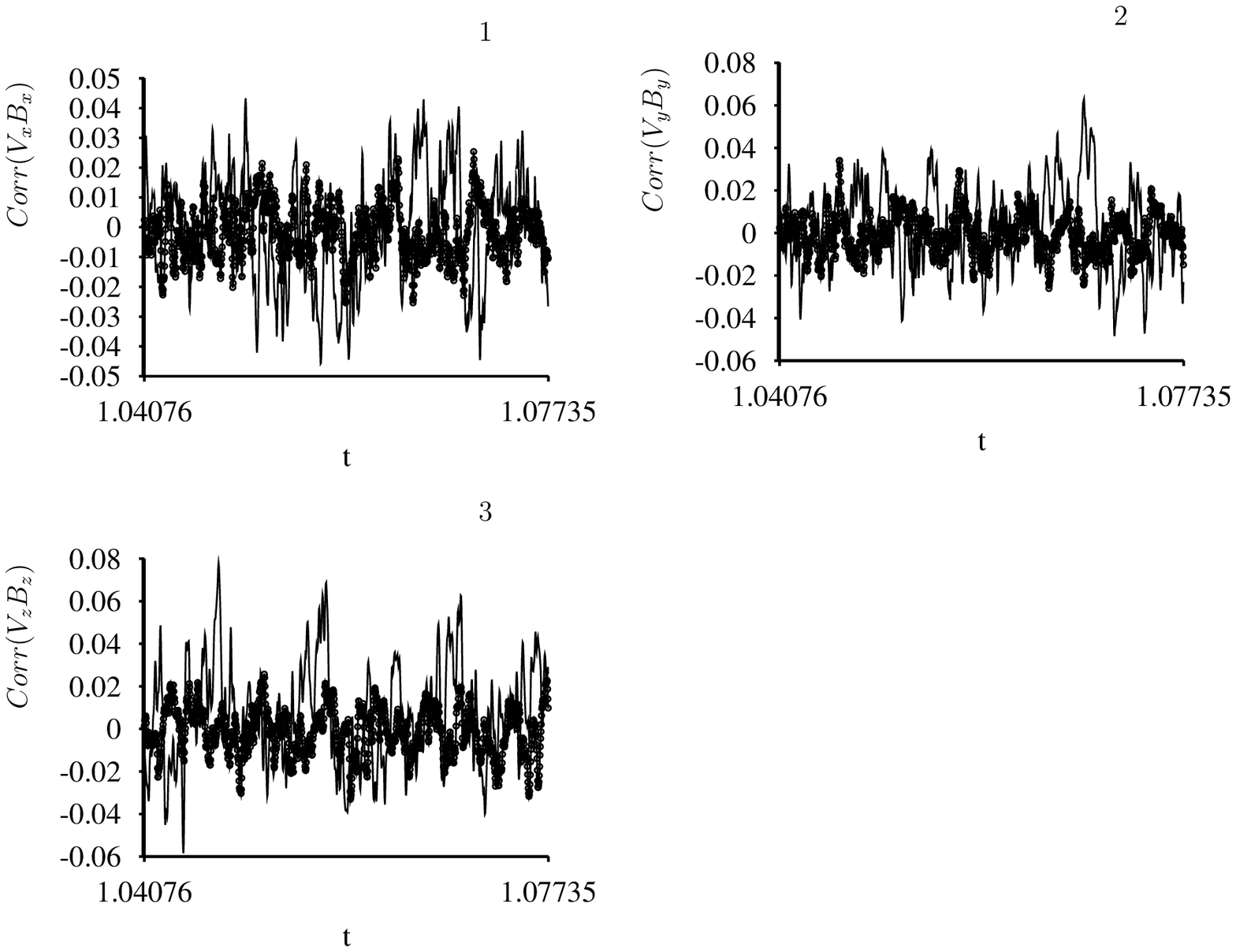,width=18cm}
\vskip -8cm
\caption{The same as in Fig.~\ref{Fig11} for regime R2.}\label{Fig12}
\end{figure*}
 
For both the regimes we observe distinct behaviours of $B^2$ and ${\widehat{B}}^2$, see 
Fig.~\ref{Fig10}. Although the same velocity field $\bf V$ was used
in both the equations for $\bf B$ and ${\widehat{\bf B}}$ , the new artificial magnetic field starts to grow exponentially. 
 We are ready to conclude that temporal synchronization of the fields in space and time is crucial for stabilization. Field  ${\bf \widehat{B}}$  is certainly less synchronized with velocity field $\bf V$ because  the Lorentz force based on ${\bf \widehat{B}}$ has been omitted. This idea is supported by our simulations of the autocorrelation functions of the magnetic and velocity fields $Corr (V_i,B_i)$,  $Corr (V_i,\widehat{B_i})$ calculated over half of the volume $z\le 0.5$, see Fig.~\ref{Fig11}, ~\ref{Fig12}.
Evidently for both the regimes R1, R2 the correlation for   ${\bf \widehat{B}}$ is less than for 
 field $\bf B$. However, the correlation is quite small for both the regimes due to the stochastic nature of the small scale fields and does not exceed a few percent,   reducing  with increasing $\Rm$. 
   The reduction of correlation for all the components corresponds to  the reduction of alignment of fields $\bf V$ and $\bf B$, when the Lorentz force is omitted. The real magnetic field $\bf B$ has stronger alignment.
  We adopt the explanation given by  \cite{CatTob2009, TBr2008} that the instability of ${\bf \widehat{B}}$ is due to the difference in the stability criteria for the full dynamo equation with quenched $\bf V_q$, and single induction equation with given $\bf V_q$.

\section{Discussion}

The transition from kinematic to saturated dynamo regime in cyclonic convection is accompanied by the reconstruction 
 of the flow as well as of the magnetic field. In general, the change of the kinetic energy is not crucial: it can even increase during the transition. More important is the reconstruction of the flow patterns. 
 The growing of the  magnetic field $\bf B$ from the quasi-stationary convective state  with non-zero kinetic helicity  $\chi^{ H}(z)$  is defined by the leading eigen solution. This first mode  grows up to the level when the maximum of the spectra at $k\sim k_c$ reaches the
  saturated level close to the equipartition value on this scale. The growth of field $\bf B$ then stops at $k_c$.    The small-scale field  produces the magnetic $\alpha^{\cal M}$-effect which suppresses the total $\alpha$-effect 
    $(\ref{sys3})$.      
    According to  (\ref{sys1}), the transition to the saturated regime requires an increase of the  large-scale magnetic field, which takes  place at diffusion time $t\sim \rm R_m$. Some change of the magnetic energy flux in the wave-space is also observed.     
     The long-term fitting to the saturated regime is also predicted by the dynamical models of $\alpha$-quenching \citep{Kleeorin1995}.     
     During this time the large-scale magnetic field still grows to the limit defined by the restriction on the magnetic helicity conservation. 
         The Lorentz force, which provides the correlation of the velocity and magnetic fields $\bf V$ and $\bf B$ and their alignment,  plays a crucial role.

\end{document}